\documentclass[doublecol]{epl2}

\usepackage{amssymb}
\usepackage{amsmath}
\usepackage{graphicx}

\title{Sphere penetration by impact in a granular medium:\\ A collisional process}

\author{A. Seguin\inst{1} \and Y. Bertho\inst{1} \and P. Gondret\inst{1} \and J. Crassous\inst{2}}
\shortauthor{A. Seguin, Y. Bertho, P. Gondret and J. Crassous}

\institute{
  \inst{1} Univ Paris-Sud, Univ Paris 6, CNRS, Lab FAST, B\^at.~502, Campus Univ, F-91405 Orsay, France\\
  \inst{2} Univ Rennes 1, Institut de Physique de Rennes (UMR UR1-CNRS 6251), B\^{a}t.~11A, Campus Beaulieu, F-35042 Rennes, France
}

\abstract{The penetration by a gravity driven impact of a solid sphere into a granular medium is studied by two-dimensional simulations. The scaling laws observed experimentally for both the final penetration depth and the stopping time with the relevant physical parameters are here recovered numerically without the consideration of any microscopic solid friction but with dissipative collisions only. Dissipative collisional processes are thus found as essential in catching the penetration dynamics in granular matter whereas microscopic frictional processes can only be considered as secondary effects.}

\begin{document}

\maketitle

\section{Introduction}
\label{sec_intro} A better knowledge of the impact of a solid object into a granular target has many applied interests from both geologist and ballistic point of view~\cite{Melosh89,Zukas90}. Since a few years, numerous studies have been carried out by physicists interested in the ejection process~\cite{thoroddsen01,Royer07,Caballero07,Marston08,Deboeuf09,Lohse04}, the crater morphology~\cite{Walsh03,Uehara03,Zheng04,Debruyn07} and the penetration dynamics~\cite{Ciamarra04,Ambroso05,Ambroso05b,Hou05,Katsuragi07,Lohse04,Tsimring05,goldman08,bourrier08}, searching for scaling laws for the crater size ~\cite{Walsh03,Uehara03,Zheng04,Debruyn07} and penetration depth~\cite{Ciamarra04,Ambroso05,Ambroso05b,Hou05,Katsuragi07,Tsimring05,goldman08,bourrier08,Newhall03,debruyn04,Seguin08}. For the ejection process, a spectacular thin granular jet raising very high can be observed after the impact on small grains of rather low packing fraction~\cite{thoroddsen01,Royer07,Caballero07,Marston08,Lohse04} whereas an opening granular corona is seen for larger grains of rather high packing fraction~\cite{Deboeuf09}: the effect of air is crucial in the former case~\cite{Royer07,Caballero07} whereas it is negligible in the latter one. For the crater morphology, the scaling laws found for high energy impacts of planetary interest~\cite{Melosh89} stand for low energy impacts of small scale laboratory experiments~\cite{Walsh03,Uehara03,Zheng04,Debruyn07}, indicating some universal physical processes involved in the crater formation. For the penetration dynamics, the observed deceleration of the impacting sphere towards its final stop is usually explained by a complex drag force resulting from frictional and collisional processes and involving several terms: a linear depth dependent term~\cite{Katsuragi07} arising from solid friction and velocity-dependent terms of linear or quadratic form arising from the ballistic~\cite{Ambroso05,Katsuragi07,Tsimring05} or the fluid mechanics point of view~\cite{debruyn04}. Such different force terms are not easy to extract from the sphere trajectory usually tracked by video means~\cite{Ciamarra04,Ambroso05,Ambroso05b,Hou05,Katsuragi07}, but recently direct force measurements with an accelerometer inside the impacting sphere~\cite{goldman08} reveal that a force proportional to the velocity squared is indeed experienced by the impacting sphere at least during its first penetration stage at high velocity and shallow depth. The stopping time of the sphere has been studied in different experimental works and displays a striking and rather counter-intuitive behaviour: it is a decreasing function of the impact velocity with an asymptotic plateau for large enough impact velocities~\cite{Ciamarra04,Katsuragi07,goldman08}. The characteristic time scaling for the plateau value was proposed to be $\tau=(d/g)^{1/2}$ \cite{Ciamarra04,Katsuragi07} or $\tau=(\rho/\rho_g)^{1/4}(d/g)^{1/2}$~\cite{goldman08} for a velocity larger than the typical characteristic velocity $V=(gd)^{1/2}$~\cite{Katsuragi07,goldman08}, where $\rho_g$ is the density of the grains, $d$ and $\rho$ are the diameter and the density of the sphere. The observed scaling law for the final penetration depth $\delta$ that may be written as $\delta /d \propto (\rho /\rho_g)^{\beta}(H/d)^\alpha$, where $H$ is the total falling distance covered from release to rest, is not yet satisfactorily explained, as well as the values of the two power exponents $\alpha$ and $\beta$ found to be around 1/2 in experimental or numerical works~\cite{Uehara03,Ambroso05,debruyn04,Seguin08,Tsimring05,bourrier08}. Experiments are essentially 3D (except the real 2D experiment of Ciamarra \etal~\cite{Ciamarra04}) whereas the numerical simulations are 2D with similar values for the power exponents. The existence of a finite penetration depth $\delta$ is always referred to the existence of a non-zero solid friction $\mu$ in contrast to the case of simple fluids where the sphere would not stop but reach a limit velocity in the absence of solid friction. The $\mu$ dependence of $\delta$ was proposed to be $\delta \propto \mu^{-1}$~\cite{Uehara03} (the penetration depth would thus be no more finite for zero solid friction) from experimental investigation varying the grain material. But experimentally, it is hard to change the coefficient of solid friction in a large range and without changing other crucial parameters such as the coefficient of restitution and the solid fraction.

In the present paper, we show by numerical simulations that no microscopic solid friction is necessary to explain a finite penetration depth for a sphere impacting on granular matter. Furthermore, we show that the scaling law $\delta /d\propto (\rho /\rho_g)^{\beta}(H/d)^\alpha$ observed experimentally or numerically with non-zero microscopic solid friction still stands in the zero microscopic solid friction limit. We also recover that the stopping time $t_s$ is constant at large enough impact velocities and show that it scales as $t_s \simeq \tau = (\rho d/\rho_g g)^{1/2}$ for impact velocities larger that the typical characteristic velocity $V=(\rho gd/\rho_g)^{1/2}$. These scalings clarify the previous scalings discussed above. The numerical results will be shown and discussed after having presented the 2D numerical method used in the present paper.

\section{Numerical method}
\label{sec:NumMed}

\begin{figure}[t]
\begin{center}
\includegraphics[width=80mm]{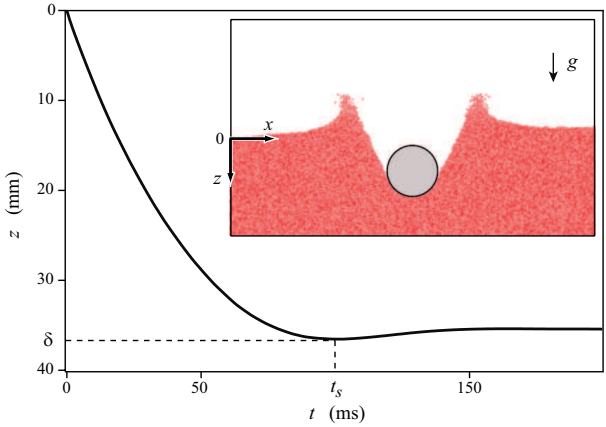}
\end{center}
\caption{Time evolution of the penetration $z$ of a sphere impacting the granular medium, where $z=0$ corresponds to the initial surface of the grain piling. The projectile of diameter $d=30\,d_g=30\un{mm}$ and of density $\rho=\rho_g=2520\un{kg\,m^{-3}}$ hits the grains at the velocity $v_i=1\un{m\,s^{-1}}$. Inset: snapshot during penetration by impact.}
\label{Fig01}
\end{figure}

We use the method of molecular dynamics to perform two-dimensional simulations in the geometry shown in the inset of fig.~\ref{Fig01}. The granular target is prepared by the sedimentation of an initial dilute configuration under the action of the gravity acceleration $g=9.8\un{m\,s^{-2}}$. The grains are modeled as a random packing of spheres of mean diameter $d_g$, mass $m_g$ and density $\rho_g$ contained in a rectangular box bounded by hard walls. In order to avoid any crystallization of the packing, a slight dispersion in the grain diameter is introduced, with an uniform distribution in the range $0.8d_g$ to $1.2d_g$. Before the impact, the packing fraction in the granular medium is $\phi\simeq 0.83$. The projectile is a sphere of diameter $d$, mass $m$ and density $\rho$ which is dropped onto the granular packing. The projectile is thrown downwards at the velocity required, and its value at impact will be noted $v_i$. This is equivalent to the usual experimental situation where the sphere is dropped from the height $h$ and impacts the granular surface with the velocity $v_i=\sqrt{2gh}$. Note that the box containing the granular medium is large enough ($> 8d$) so that the projectile is not affected by the surrounding walls and the layer of grains is high enough to avoid any bottom wall effects during the penetration \cite{Seguin08}. The number of grains in the simulations ranges from $10^{4}$ for small boxes to $10^{5}$ for largest ones.

As the goal of this paper is to show that microscopic solid friction is not essential in explaining the finite penetration of a projectile into a granular material, we do not take into account any static nor dynamic friction between the grains. The interaction forces are thus taken as purely normal with no tangential components. The interaction force between two grains, or between one grain and the projectile or the bounded walls, is modeled as a dissipative Hertz law \cite{Schafer96} such as
\begin{equation}
F_n= -k\xi^{3/2} - \gamma \frac{\mathrm{d}\xi}{\mathrm{d}t},
\label{eq1.1}
\end{equation}
where $k$ is the non-linear stiffness, $\gamma$ is a damping coefficient, and $\xi$ and $d\xi/dt$ are respectively the interpenetration and the velocity of interpenetration of the grains. For two identical spherical grains of diameter $d_g$, Young modulus $E$ and Poisson coefficient $\nu$, the non-linear stiffness $k$ is given by $k=2E\sqrt{2d_g}/3(1-\nu^2)$. The collision time for a non-dissipative contact ($\gamma=0$) between two grains is
\begin{equation}
\tau_c \approx 3.21 \left (\frac{m_\mathrm{eff}}{k} \right )v_n^{-1/5},
\label{eq1.2}
\end{equation}
where $v_n$ is the relative normal velocity and $m_\mathrm{eff}=m_g/2$ the effective mass for two identical colliding grains. The collision time is not very different in the case of a non-zero dissipation ($\gamma \neq 0$). The equations of motion for the grains are integrated using a standard second order Verlet algorithm. The choice for the numerical time step $\Delta t$ must be such that $\Delta t \ll \tau_c$ in order to ensure numerical accuracy.

In the following, we present numerical simulations for a granular material composed of glass spheres (density $\rho_g=2520\un{kg\,m^{-3}}$) with an effective elastic modulus $E^*=E/(1-\nu^2)=69\times10^9\un{Pa}$ and of mean diameter $d_g=1\un{mm}$ (mass $m_g=1.3\un{mg}$). The non-linear stiffness is thus $k=2\times 10^9\un{N\,m^{3/2}}$ and the collision time for a typical collision velocity $v_n=1\un{m\,s^{-1}}$ is $\tau_c=2.7\un{\mu s}$. The time step is chosen as $\Delta t=0.1\un{\mu s}\ll\tau_c$. For a non-zero damping coefficient $\gamma$, the coefficient of normal restitution for normal incidence $e_n=-v_n^f/v_n^i$, which is the ratio of the relative normal velocity after an impact over the velocity before the impact, is smaller than the ideal limit value 1 for perfect elastic collisions. More precisely, with model eq.~(\ref{eq1.1}), $e_n$ decreases slowly with the collision velocity as $(1-e_n) \propto (v_n^i)^{-1/5}$ \cite{Schafer96}. With the damping value $\gamma=0.065\un{kg\,s^{-1}}$, the restitution coefficient is $e_n=0.9$ for $v_n^i=1\un{m\,s^{-1}}$. It should be stressed that the following results are qualitatively only weakly dependent on the damping factor value.

\section{Numerical results and discussion}
\label{sec:Results}

Figure~\ref{Fig01} displays the time evolution of the position $z(t)$ of the projectile during its penetration through the grains, where $z$ is the distance between the initial horizontal free surface of the granular medium and the bottom of the impacting sphere. The penetration increases with time up to a maximum depth $\delta$ at the time $t_s$. Note that a small rise of the sphere, of typically a few percent of the total penetration, is observed at the end of most of the runs. We attribute this effect to the absence of microscopic static friction in the interaction law [eq.~(\ref{eq1.1})]. Indeed, the grains ejected during the collision process are redeposited on
the granular material and exert an increasing pressure on it, inducing a downwards motion of the granular material. This small effect was not reported in numerical simulations including microscopic solid friction~\cite{Tsimring05}. In the following, we shall consider this maximum penetration $\delta$ as the final penetration depth and the corresponding time $t_s$ as the stopping time. Note that dropping the sphere from slightly different $x$-positions gives very similar $z(t)$ even if the acceleration signals are quite different.

\subsection{Penetration depth}

\begin{figure}[t]
\begin{center}
\includegraphics[width=80mm]{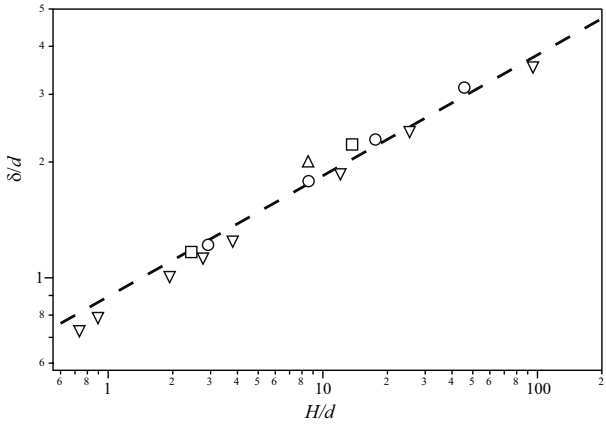}
\end{center}
\caption{Normalized penetration depth $\delta/d$ of the sphere as a function of the normalized total falling distance $H/d$ for different projectile diameters: $d=20d_g$~($\triangledown$),
$d=30d_g$~($\circ$), $d=40d_g$~($\square$) and $d=70d_g$~($\triangle$). The sphere/grains density ratio is $\rho/\rho_g=1$. (--~--)~Power law fit $\delta /d\propto (H/d)^\alpha$, with $\alpha\simeq 0.31$.}
\label{Fig02}
\end{figure}

\begin{figure}[t]
\begin{center}
\includegraphics[width=80mm]{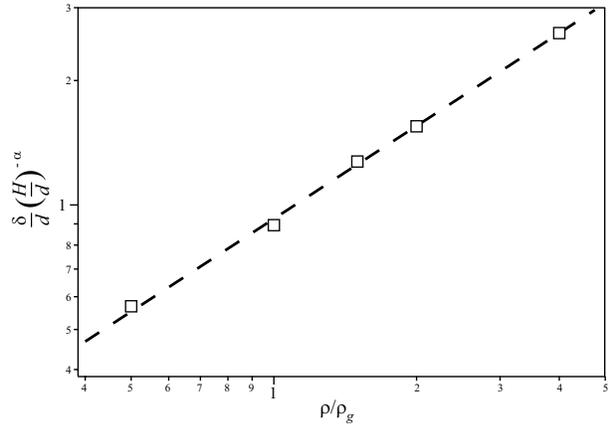}
\end{center}
\caption{$(\delta/d)(H/d)^{-\alpha}$ as a function of the
sphere/grains density ratio $\rho/\rho_g$. (--~--)~Power law fit
$(\delta/d)(H/d)^{-\alpha} \propto (\rho/\rho_g)^\beta$, with
$\beta\simeq 0.74$.} \label{Fig03}
\end{figure}

Figure~\ref{Fig02} shows the evolution of the normalized penetration depth $\delta/d$ as a function of the normalized total falling distance $H/d$, where $H=h+\delta$ is the sum of the free-fall height $h=v_i^2/2g$ and the penetration depth. The diameter of the impacting sphere ranges from $d=20d_g$ to $70d_g$, and its density is here kept constant and equalled to the grain density ($\rho=\rho_g$). In the log-log plot of fig.~\ref{Fig02}, the data are aligned along a straight line of slope $\alpha=0.31 \pm 0.02$ meaning that the penetration depth $\delta$ varies with $H$ following a power law $\delta/d \propto (H/d)^\alpha$. The variation of the penetration depth with the density (\emph{i.e.} with the mass) of the projectile is shown in fig.~\ref{Fig03}, where $(\delta /d)/(H/d)^{-\alpha}$ is plotted as a function of the density ratio $\rho /\rho_g$ between the projectile and the grains. The penetration depth is found to depend on $\rho /\rho_g$ with a power law of the form $\delta /d\propto (\rho /\rho_g)^\beta$, with $\beta=0.74 \pm 0.02$. Finally, grouping the fall height and the mass dependencies, one obtains
\begin{equation}
\frac{\delta}{d} =A \left (\frac{\rho}{\rho_g}\right )^\beta \left (\frac{H}{d}\right )^\alpha,
\label{eq2.1}
\end{equation}
with $A =0.92 \pm 0.02$. This power law scaling for the penetration depth is observed in various experimental and numerical studies~\cite{Uehara03,Ambroso05,debruyn04,Seguin08,Tsimring05,bourrier08}. The value of the exponent $\alpha\simeq 0.31$ is close to the commonly reported values between $0.33$ and $0.40$, and the exponent $\beta\simeq 0.74$ characterizing the dependence with the density ratio is not far from the reported values ranging from $0.50$ to $0.61$. Thus, the impacting sphere stops at a finite depth without any microscopic friction and the scaling laws for the penetration depth are in agreement with those observed experimentally or numerically with friction.

\subsection{Forces}

Let us now examine the forces undergone by the sphere after the impact, to extract the main ingredients leading to its stop. The force exerted by the grains on the sphere depends mainly on the penetration depth $z$ and on the velocity $v$ of the projectile~\cite{Katsuragi07,Tsimring05,goldman08}. For dense packings, experimental results show that this resistance force may be separated into two independent functions of position $F_z(z)$ and velocity $F_v(v)$ so that the Newton law for the sphere motion can be written as
\begin{equation}
\frac{\mathrm{d}v}{\mathrm{d}t}= g- \frac{F_z(z)}{m}- \frac{F_v(v)}{m}.
\label{eq4.1}
\end{equation}
$F_z$ is usually attributed to solid friction~\cite{Katsuragi07,Lohse04} and $F_v$ is considered of collisional or inertial origin~\cite{Katsuragi07,Tsimring05,goldman08}.
\begin{figure}[t!]
\begin{center}
\includegraphics[width=80mm]{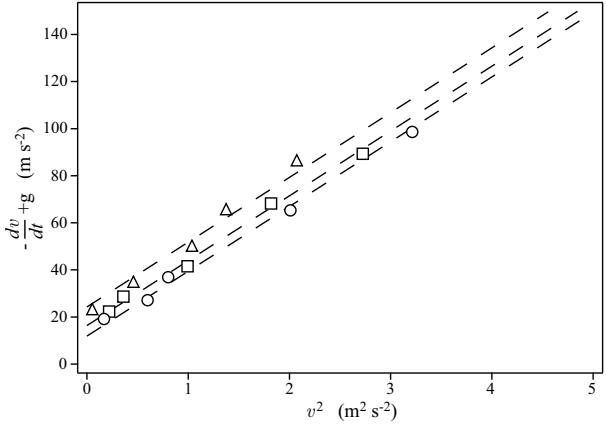}
\end{center}
\caption{Reduced acceleration $-\mathrm{d}v/\mathrm{d}t+g$ of the impacting sphere
(diameter $d=20d_g$, density $\rho=\rho_g$) as a function of its
velocity squared $v^2$, at a specific depth:
($\circ$)~$z=30\un{mm}$, ($\square$)~$z=40\un{mm}$ and
($\triangle$)~$z=50\un{mm}$. (--~--)~Linear trends of the data by
$-\mathrm{d}v/\mathrm{d}t+g = F_z(z)/m + v^2/d_1$, with $d_1\simeq 36.4\un{mm}$.}
\label{Fig04}
\end{figure}

\begin{figure}[t!]
\begin{center}
\includegraphics[width=80mm]{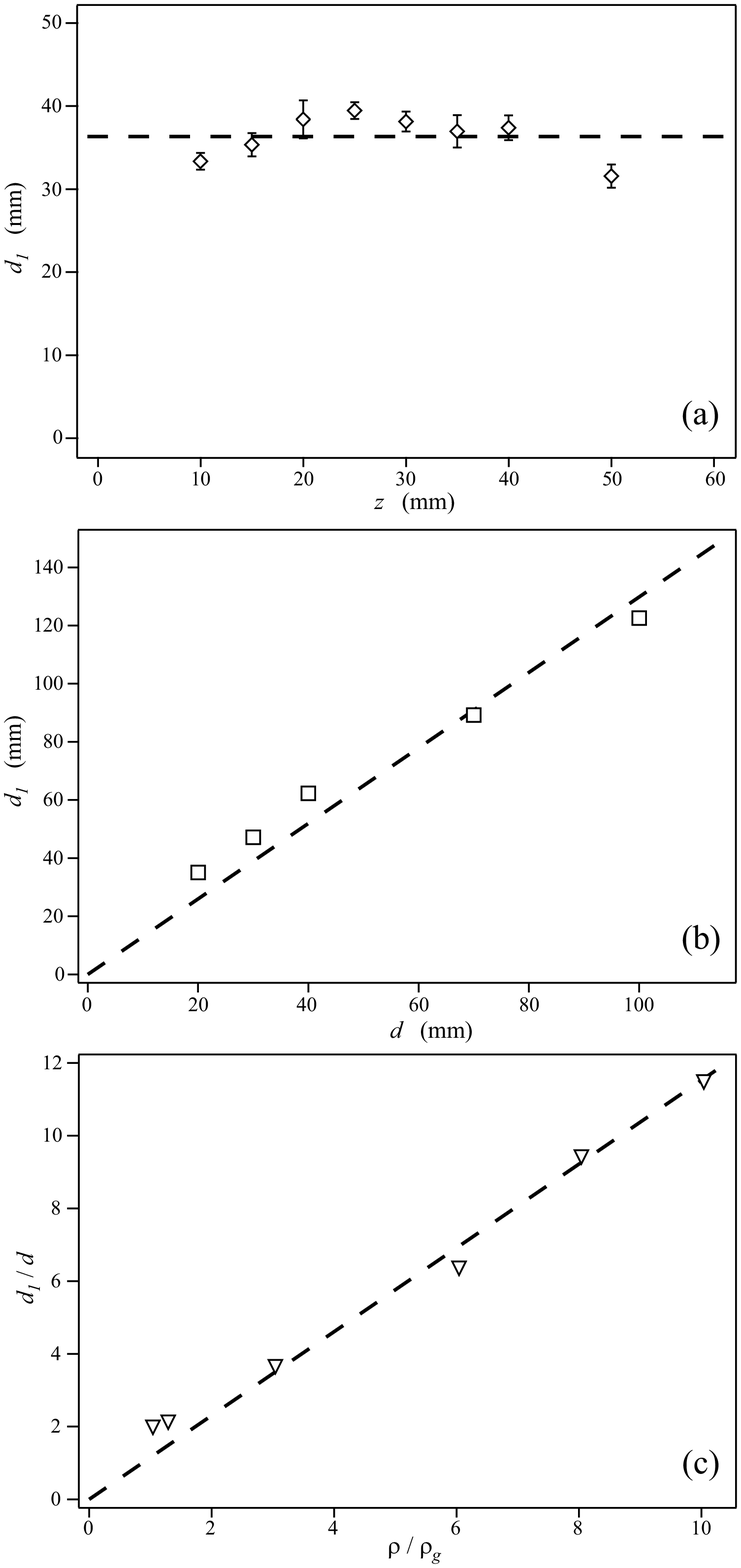}
\end{center}
\caption{(a)~Characteristic length $d_1$ as a function of the depth
$z$ of the projectile. (--~--)~Constant fit $d_1= 36.4\un{mm}$.
(b)~Characteristic length $d_1$ as a function of the diameter $d$ of
the projectile (density $\rho=\rho_g$) at a depth $z=20\un{mm}$.
(-- --)~Linear fit $d_1=1.3d$. (c)~Characteristic length $d_1/d$ as
a function of the density ratio $\rho/\rho_g$ at $z=30\un{mm}$ for a
projectile of diameter $d=20d_g$. (-- --)~Linear fit
$d_1/d=1.2\rho/\rho_g$.} \label{Fig05}
\end{figure}

Figure~\ref{Fig04} displays the reduced acceleration of the sphere as a function of its velocity at a given penetration depth ($z=30\un{mm}$) and suggests that the drag force $F_v$ is proportional to $v^2$, in agreement with the expression $mv^2/d_1$ proposed recently by Katsuragi and Durian~\cite{Katsuragi07}, where $d_1$ is a characteristic dissipative length. Note that the reduced force is non-zero at vanishing velocity as there is still a non-zero depth dependent force term $F_z(z)$. This behavior is observed with different sphere diameters in the range $20\un{mm} <d<100\un{mm}$ and different density ratio in the range $1 < \rho /\rho_g < 10$. Figure~\ref{Fig05}a shows that $d_1$ does not depend on the depth of the projectile $z$ and thus $F_v$ does not depend on $z$ as found experimentally~\cite{Katsuragi07}, which justify the writing of eq.~(\ref{eq4.1}). Figures~\ref{Fig05}b and \ref{Fig05}c show that $d_1$ is proportional both to the projectile diameter $d$ and to the density ratio $\rho/\rho_g$ leading to $d_1\propto \rho d/\rho_g$. The velocity dependent force term $F_v(v)$ thus scales as $F_v/m \propto \rho_g v^2 / \rho d$ indicating its inertial or collisional origin.

\begin{figure}[t!]
\begin{center}
\includegraphics[width=80mm]{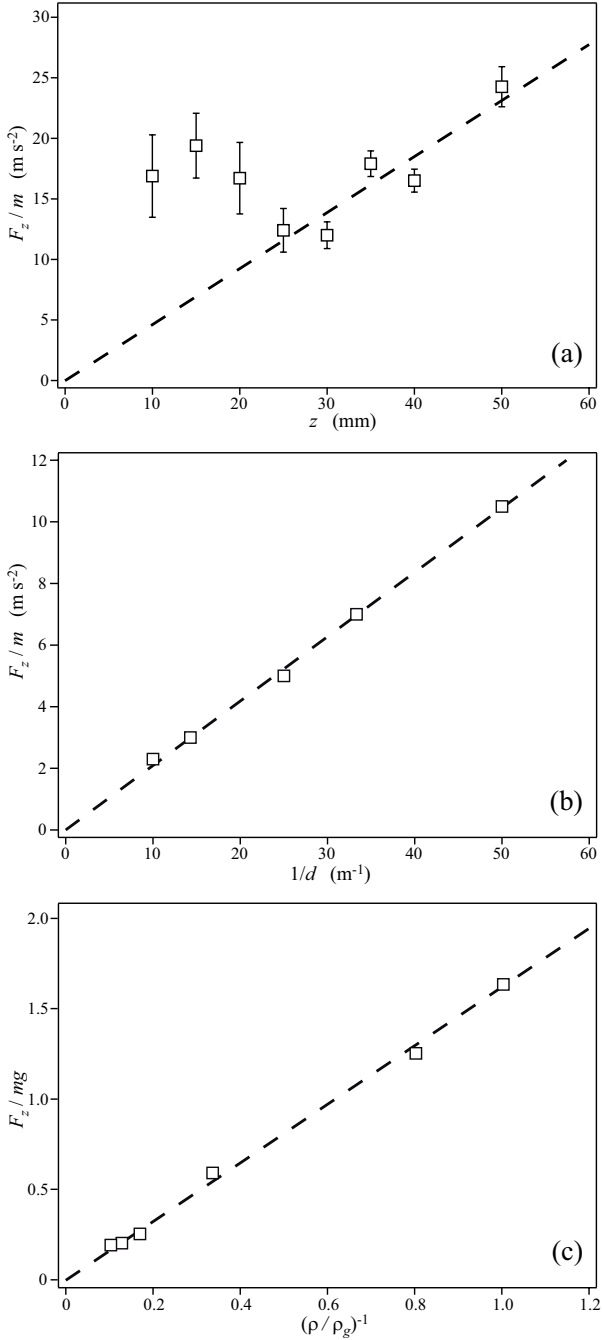}
\end{center}
\caption{(a)~$F_z/m$ as a function of the depth $z$ of the
projectile (diameter $d=20d_g$, density $\rho=\rho_g$); (b)~$F_z/m$
as a function of the inverse of the diameter $1/d$ of the projectile
(density $\rho=\rho_g$) at a depth $z=20\un{mm}$; (c)~$F_z/mg$ as
a function of the inverse of the density ratio $(\rho/\rho_g)^{-1}$
at $z=30$\,mm for a projectile of diameter $d=20d_g$. (--~--)~Linear
fits of numerical data.} \label{Fig06}
\end{figure}

Plotting now in fig.~\ref{Fig06}a the force term at vanishing velocity $v$ as a function of $z$ shows that the simple dependence $F_z\propto z$ proposed by Katsuragi and Durian~\cite{Katsuragi07} is compatible with our data despite the high scattering
at low $z$ ($z\lesssim d$) where the sphere is not fully immersed in the granular medium. Figures~\ref{Fig06}b and \ref{Fig06}c show that $F_z/m$ is proportional to $1/d$ and $\rho_g/\rho$. The depth dependent force term $F_z$ thus scales as $F_z(z)/m \propto\rho_ggz / \rho d$. This force term linear in $z$ has been previously proposed and seen by various authors \cite{Katsuragi07,Tsimring05,goldman08} with a solid friction origin, but the extracted coefficient of friction necessary to fit the data was far from the standard values \cite{Katsuragi07}. Here, the $F_z(z)$ force term does not come from microscopic solid friction as there is no microscopic solid friction in our numerical simulations. The depth dependent term $F_z(z)$ can thus simply viewed as a hydrostatic term as already suggested in \cite{stone04} for the experimental penetration of flat plates and in \cite{Lohse04} for an impacting sphere in very loose sand. The pressure increases linearly with the depth as $\rho_g g z$ and so is the force $F_z$ on the sphere. This ``hydrostatic'' force term is not Archimedean as the penetrating sphere is never immersed in the granular packing before it stops as can be seen here numerically (see the inset snapshot of fig.~\ref{Fig01}) and experimentally \cite{goldman08}.
It follows that the Newton's law for the projectile during its penetration may be written as
\begin{equation}
\frac{\mathrm{d}v}{\mathrm{d}t}=g-a_z\frac{\rho_g g}{\rho d}z - a_v\frac{\rho_g v^2}{\rho d},
\label{eq4.2}
\end{equation}
where $a_z=1\pm 0.1$ and $a_v=0.8\pm 0.1$ are numerical prefactors.

\subsection{Stopping time}
\begin{figure}[t!]
\begin{center}
\includegraphics[width=80mm]{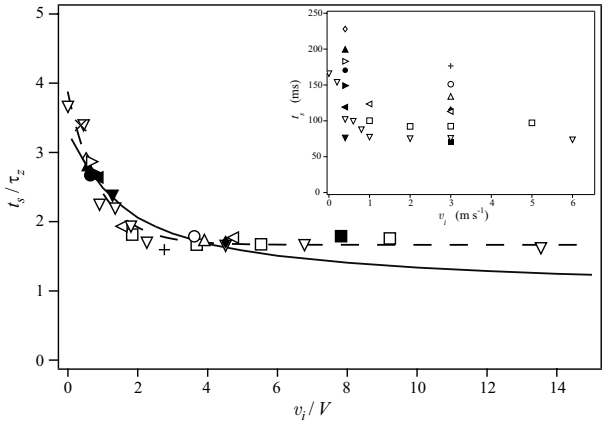}
\end{center}
\caption{Normalized stopping time $t_s/\tau$ as a function of the
normalized impact velocity $v_i/V$, for: $\rho/\rho_g=0.5$ and
($\blacktriangledown$)~$d=20\un{mm}$,
($\blacksquare$)~$d=30\un{mm}$,
($\blacktriangleleft$)~$d=40\un{mm}$,
($\blacktriangleright$)~$d=60\un{mm}$, ($\bullet$)~$d=80\un{mm}$,
($\blacktriangle$)~$d=100\un{mm}$; $\rho/\rho_g=1$ and
($\triangledown$)~$d=20\un{mm}$, ($\square$)~$d=30\un{mm}$,
($\triangleleft$)~$d=40\un{mm}$, ($\circ$)~$d=70\un{mm}$;
$\rho/\rho_g=1.5$ and ($\blacklozenge$)~$d=30\un{mm}$;
$\rho/\rho_g=2$ and ($\triangleright$)~$d=20\un{mm}$,
($\triangle$)~$d=30\un{mm}$; $\rho/\rho_g=3$ and
($\diamond$)~$d=20\un{mm}$; $\rho/\rho_g=4$ and ($+$)~$d=30\un{mm}$;
$\rho/\rho_g=6$ and ($\times$)~$d=20\un{mm}$. (-- --)~Guideline for the eyes.
(---)~Calculated stopping time from model eq.~(\ref{eq4.2}).
Inset: stopping time $t_s$ as a function of the impact velocity
$v_i$ for the same set of data.} \label{Fig07}
\end{figure}

Let us now focus on the stopping time. The inset of fig.~\ref{Fig07}
displays the stopping time $t_s$ of the projectile as a function of
its impact velocity $v_i$. Numerical data are observed to be very
scattered depending on the velocity, diameter and density of the
projectile. More precisely, $t_s$ decreases with the impact velocity
$v_i$ for any given size and density of the projectile and increases
with the projectile diameter $d$ and with its density ratio $\rho
/\rho_g$. Two time scales can be extracted from the non-linear
differential equation~(\ref{eq4.2}) by considering independently the
two force terms. Considering only the depth dependent force term,
the characteristic time $\tau_z$ for the stopping time is
\begin{equation}
\tau_z = \left (\frac{\rho d}{\rho_g g}\right )^{1/2},
\end{equation}
which is independent of the impact velocity. Considering now only
the non-linear velocity dependent force term in eq.~(\ref{eq4.2})
leads to the characteristic time
\begin{equation}
\tau_v = \frac{\rho d}{\rho_g v_i},
\end{equation}
which depends on the impact velocity $v_i$. As both force terms play
a non-negligible role in the penetration (the velocity dependent
force term decreases from its maximal value at impact to zero at the
stop whereas the depth dependent force term increases from zero at
impact to its maximal value at the stop), the total stopping time
$t_s(\tau_z, \tau_v)$ is a combination of the two characteristic
time scales $\tau_z$ and $\tau_v$, and can thus be expressed as
$t_s/\tau_z = f(\tau_z/\tau_v)$. Note that the time scale ratio is
$\tau_z/\tau_v = v_i (\rho_g/\rho g d)^{1/2}$, which corresponds
also to the velocity ratio $v_i/V$ of the impact velocity $v_i$ to
the characteristic velocity $V=(\rho gd/\rho_g)^{1/2}$.

By using the characteristic time scale $\tau_z$ and velocity scale $V$, \emph{i.e.} by plotting $t_s/\tau_z$ as a function of $v_i/V$, all the data collapse on the same master curve with two distinct parts (fig.~\ref{Fig07}): (\emph{i}) for low enough impact velocities ($v_i/V \lesssim 2$), the stopping time decreases with increasing impact velocity; (\emph{ii}) for large enough impact velocities ($v_i/V \gtrsim 2$), a plateau is reached, so that the stopping time does not depend on the impact velocity and tends to the constant value $t_s/\tau_z \simeq 1.7$. The existence of a single curve in the normalized plot $t_s/\tau_z$ versus $v_i/V$, and the critical values $v_i/V \simeq 2$ and $t_s/\tau_z \simeq 1.7$ close to 1 indicate that the typical velocity scale $V$ and time scale $\tau_z$ are relevant physical parameters for the problem, which validate the two independent model forces acting on the penetrating sphere. The stopping time calculated by eq.~(\ref{eq4.2}) with a stop condition at $v=0$ agrees with the simulation data to within 20\% (fig.~\ref{Fig07}). The calculated penetration depth $\delta/d$ obtained by solving eq.~(\ref{eq4.2}) deviates from the data by as much as 50\% since small errors in the force terms in the approximated model equation have a more important effect on the integrated position than on the stopping time.

\section{Conclusion}
By performing numerical simulations for an impacting sphere in a frictionless granular material, one obtains the same scaling law for the penetration depth as in simulations with solid friction or real experiments. This shows that dissipation by microscopic solid friction can be ignored and dissipation by collisions is sufficient to catch the penetration dynamics in granular matter. Analysing the forces reveals that besides a velocity squared force term from collisional origin, exists a depth dependent force term. This pressure increasing term with the depth is sufficient to stop the sphere, and steric hindrance with dissipation prevents the sphere from settling, by contrast to simple liquids. The scalings found for the two force terms allow for the prediction of the scaling of the stopping time which is indeed observed: a plateau value at high impact velocity and an increasing value at decreasing velocity. This scaling for the stopping time without any microscopic solid friction is also in agreement with what was observed previously. If microscopic solid friction appears needless to explain the scaling laws for both the final penetration depth and the stopping time of the projectile, it would certainly affect their precise value from a quantitative point of view.

\acknowledgments{The authors thank John Hinch and Marc Rabaud for fruitful discussions.}


\end{document}